\documentclass[onecolumn]{aa}
\usepackage{graphicx}
\usepackage{txfonts}
\begin{document}
\def\gtsim {>\kern-1.2em\lower1.1ex\hbox{$\sim$}}
\def\ltsim {<\kern-1.2em\lower1.1ex\hbox{$\sim$}}
\title{Nonradial Oscillations of Neutron Stars with a Solid Crust}

\subtitle{Analysis in the Relativistic Cowling Approximation}

\author{
Shijun Yoshida\inst{1}
\and
Umin Lee\inst{2}
}

\offprints{S. Yoshida}

\institute{Centro Multidisciplinar de Astrof\'{\i}sica -- CENTRA,
           Departamento de F\'{\i}sica, Instituto Superior T\'ecnico, 
           Av. Rovisco Pais 1, 1049-001 Lisboa, Portugal \\
           \email{yoshida@fisica.ist.utl.pt; yoshida@astr.tohoku.ac.jp}
\and
           Astronomical Institute, Graduate School of Science,
           Tohoku University, Sendai 980-8578, Japan\\
           \email{lee@astr.tohoku.ac.jp}
}

\date{Received ................, Accepted ....................}

\abstract{
Nonradial oscillations of relativistic neutron stars with a solid 
crust are computed in the relativistic Cowling approximation, in 
which all metric perturbations are ignored. For the modal analysis, 
we employ three-component relativistic neutron star models with a 
solid crust, a fluid core, and a fluid ocean. As a measure for the 
relativistic effects on the oscillation modes, we calculate the 
relative frequency difference defined as 
$\delta\sigma/\sigma\equiv(\sigma_{GR}-\sigma_N)/\sigma_{GR}$, where 
$\sigma_{GR}$ and $\sigma_R$ are, respectively, the relativistic and 
the Newtonian oscillation frequencies. The relative difference 
$\delta\sigma/\sigma$ takes various values for different oscillation 
modes of the neutron star model, and the value of $\delta\sigma/\sigma$ 
for a given mode depends on the physical properties of the models.
We find that $|\delta\sigma/\sigma|$ is less than $\sim0.1$ for most of 
the oscillation modes we calculate, although there are a few exceptions
such as the fundamental (nodeless) toroidal torsional modes in the crust,
the surface gravity modes confined in the surface ocean, and the core 
gravity modes trapped in the fluid core. We also find that the modal 
properties, represented by the eigenfunctions, are not strongly affected 
by introducing general relativity. It is however shown that the mode 
characters of the two interfacial modes, associated with the core/crust 
and crust/ocean interfaces, have been interchanged between the two
through an avoided crossing when we move from Newtonian dynamics to 
general relativistic dynamics.
}

\authorrunning{S. Yoshida \& U. Lee}
\titlerunning{Nonradial Oscillations of Neutron Stars}

\maketitle

\section{Introduction}

After the discovery of the $r$ mode instability in neutron stars
driven by gravitational
radiation reaction (Andersson 1998; Friedman \& Morsink 1998), nonradial
oscillations of relativistic neutron stars has attracted much wider interest
than before in astrophysics
(see a recent review, e.g., by Andersson \& Kokkotas 2001).
The relativistic formulation of nonradial oscillations of $fluid$ neutron stars was first
given by Thorne and his collaborators (e.g., Thorne \& Campolattaro 1967;
Price \& Thorne 1969; Thorne 1969a,b), and later on extended to the case of
neutron stars with a solid crust in their interior
(Schumaker \& Thorne 1983, see also Finn 1990).
Since one of the main concerns of these studies was gravitational waves
generated by the stellar pulsations (e.g., Thorne 1969a),
it was essential to include the perturbations in the metric $g_{\alpha\beta}$
in order to obtain a consistent description of the gravitational waves.
However, it was the metric perturbations that made it extremely difficult to
treat both analytically and numerically the oscillation modes in relativistic
stars.

McDermott, van Horn, \& Scholl (1983) introduced a relativistic version
of the Cowling approximation, in which all the Eulerian metric perturbations
$\delta g_{\alpha\beta}$ are neglected
in the relativistic oscillation equations derived by Thorne \& Campolattaro (1967).
Under the relativistic Cowling approximation,
they calculated the $p$-, $f$-, and $g$- modes
of relativistic $fluid$ neutron star models to examine how the
oscillation modes depend on the model properties.
The relativistic
Cowling approximation employed by McDermott, van Horn, \& Scholl (1983) was shown to be
good enough to calculate the $p$-modes for non-rotating polytropic stars
by Lindblom \& Splinter (1990), who compared the fully relativistic
oscillation frequencies to those obtained in the relativistic Cowling approximation.
Assuming slow rotation,
Yoshida \& Kojima (1997) have also carried out similar computations
for the $f$- and $p$-modes of polytropic models and they confirmed
the good applicability of the relativistic Cowling approximation.
Quite recently, Yoshida \& Lee (2002) have shown in the relativistic Cowling approximation 
the existence of the relativistic $r$ modes\footnote{We should note that lots of studies 
about the general relativistic $r$-mode have been done so far (e.g., Kojima 1998; 
Kojima \& Hosonuma 2000; Lockitch, Andersson \& Friedman 2001;   
Yoshida 2001; Ruoff \& Kokkotas 2001).} 
with $l=m$, which are regarded as a counter part
of the Newtonian $r$ modes.
We believe that the relativistic Cowling approximation is quite useful to investigate
the oscillation modes of relativistic stars, although we understand that
under the approximation we cannot
discuss inherently relativistic oscillation modes like $w$-modes
(see, e.g., Andersson, Kojima, \& Kokkotas 1996).

In their Newtonian calculations, McDermott, van Horn, \& Hansen (1988) have shown that
cold neutron stars with a solid crust can support a rich variety of oscillation modes
(see also Strohmayer 1991; Lee \& Strohmayer 1996; Yoshida \& Lee 2001).
The purposes of this paper are to calculate
the various oscillation modes of relativistic neutron stars with a solid crust in the
relativistic Cowling approximation, and to discuss the effects of general relativity
on the modal property of the oscillation modes.
We regard this paper as an extension of the studies by McDermott, van Horn, \&
Scholl (1983) and McDermott, van Horn, \& Hansen (1988).
In \S 2, we derive relativistic oscillation equations for the solid crust
in the relativistic Cowling approximation
following the formulation developed by
Schumaker \& Thorne (1983) and Finn (1990).
Numerical results are discussed in \S 3 for nonradial modes of
three relativistic neutron star models with a solid crust.
\S 4 is devoted to discussion and conclusion.
In this paper, we use units in which $c=G=1$, where $c$ and $G$ denote the
velocity of light and the gravitational constant, respectively.

\section{Formulation}

For modal analysis of neutron stars with a solid crust, we solve general relativistic
pulsation equations derived under the relativistic Cowling approximation, in which
all the metric perturbations in the matter equations are neglected.
We assume that the solid crust is in a strain-free state in the equilibrium
unperturbed state
and the strain in the crust are generated by
small amplitude perturbations superposed on the unperturbed state
(see, e.g., Aki \& Richards 1980).
For the background unperturbed state,
it is therefore possible to assume a static and spherical symmetric state,
for which the geometry is given by the following line element:
\begin{eqnarray}
(ds^2)_{(0)} = g_{(0) \alpha\beta}dx^\alpha dx^\beta =
     - e^{2 \nu(r)} dt^2 + e^{2 \lambda(r)} dr^2 + r^2 d\theta^2 +
           r^2 \sin^2 \theta d\varphi^2 \, ,
\label{metric}
\end{eqnarray}
where the subscript $(0)$ is used to refer to the unperturbed state.
The four-velocity of the stellar material in the equilibrium is given by
\begin{equation}
u_{(0)}^\alpha = e^{-\nu(r)} \, t^\alpha \, ,
\end{equation}
where $t^\alpha$ denotes the timelike Killing vector of the unperturbed
spacetime.

In the relativistic Cowling approximation, the basic equations
for pulsations are obtained from the energy and momentum conservation laws:
\begin{eqnarray}
&&u^\alpha \nabla_\beta T^\beta_\alpha = 0 \, , \ \ \ \
{\rm (energy\ conservation\ law)}
\label{ena-eq} \\
&&q^\alpha_\gamma \nabla_\beta T^\beta_\alpha = 0 \, , \ \ \ \
{\rm (momentum\ conservation\ law)}
\label{mom-eq}
\end{eqnarray}
where $\nabla_\alpha$ is the covariant derivative associated with the metric,
$T^\alpha_\beta$ is the energy-momentum tensor, and $q^\alpha_\beta$ is
the projection tensor with respect to the fluid four-velocity $u^\alpha$,
which is defined by
\begin{eqnarray}
q^\alpha_\beta = \delta^\alpha_\beta + u^\alpha u_\beta \, ,
\end{eqnarray}
where $\delta_\alpha^\beta$ denotes the Kronecker delta.

For relativistic analysis of vibration of the solid crust in neutron stars,
we follow the formulation given by Schumaker \& Thorne (1983) and Finn (1990).
If we define the rate of shear $\sigma_{\alpha\beta}$ and the
rate of expansion $\theta$ as
\begin{equation}
\sigma_{\alpha\beta}=\frac{1}{2}\, q_\alpha^\gamma q_\beta^\delta \,
(\nabla_\gamma u_\delta+\nabla_\delta u_\gamma) -
\frac{1}{3}\, q_{\alpha\beta} \, \theta \, ,
\end{equation}
\begin{equation}
\theta =  q_\alpha^\beta \nabla_\beta u^\alpha = \nabla_\alpha u^\alpha \, ,
\end{equation}
the shear strain tensor $\Sigma_{\alpha\beta}$ is determined, in terms of
$\sigma_{\alpha\beta}$ and $\theta$, as a solution of
the differential equation given by
\begin{equation}
L_u\, \Sigma_{\alpha\beta} = \frac{2}{3} \, \theta \, \Sigma_{\alpha\beta} +
\sigma_{\alpha\beta} \, ,
\label{def-Sigma}
\end{equation}
where $L_u$ is the Lie derivative along the four-velocity field of the matter
(Carter \& Quintana 1972).
Once equation (\ref{def-Sigma}) is solved to give the shear strain tensor
$\Sigma_{\alpha\beta}$ in terms of $\sigma_{\alpha\beta}$ and $\theta$
(see Schumaker \& Thorne 1983, Finn 1990),
the total stress-energy tensor is given by
\begin{equation}
T_{\alpha\beta}=\rho \,u_\alpha u_\beta + p \, q_{\alpha\beta} -2 \mu\,
\Sigma_{\alpha\beta} \, ,
\label{EMT}
\end{equation}
where $\mu$ stands for the isotropic shear modulus, and
we have assumed a Hookean relationship between the shear strain and
stress tensors. Here, $\rho$ and $p$ mean the mass-energy density and the
isotropic pressure, respectively.  Note that $\Sigma_{(0)\alpha\beta}=0$
since we have assumed the strain-free state in the equilibrium of the star.

To discuss nonradial oscillations of a star, we employ a Lagrangian perturbation
formalism (see, e.g., Friedman \& Schutz 1975; Friedman 1978), in which
a Lagrangian displacement vector is introduced to connect fluid elements in
the equilibrium state to the corresponding elements in the perturbed state.
In this formalism, the Lagrangian change $\Delta Q$ in a quantity is related to
its Eulerian change $\delta Q$ by
\begin{eqnarray}
\Delta Q = \delta Q + L_\zeta Q \, ,
\end{eqnarray}
where $L_\zeta$ denotes the Lie derivative along the displacement vector $\zeta^\alpha$.
If we apply this formalism to oscillations of relativistic stars, we have, for example,
\begin{eqnarray}
\Delta g_{\alpha\beta}=\delta g_{\alpha\beta}
+\nabla_{\alpha}\zeta_{\beta}+\nabla_{\beta}\zeta_{\alpha},
\end{eqnarray}
and
\begin{eqnarray}
\Delta u^\alpha={1\over 2}u^\alpha u^\beta u^\gamma\Delta g_{\beta\gamma}.
\end{eqnarray}
The Eulerian perturbation in the velocity field
$\delta\hat{u}^\beta \equiv{q_{(0)}}^\alpha_\beta \delta u^\beta$
is then given by
\begin{eqnarray}
\delta \hat{u}^{\alpha} =
{q_{(0)}}^\alpha_\beta (L_{u_{(0)}} \zeta)^\beta \, .
\end{eqnarray}
Notice that $\delta g_{\alpha\beta}=0$ and hence
$\delta \hat{u}^{\alpha} = \delta u^\alpha$
in the relativistic Cowling approximation.
Because the background state is in a hydrostatic equilibrium,
the time dependence of all the perturbed quantities can be
given by $e^{i\sigma t}$, where $\sigma$ is a constant
frequency measured by an inertial observer at the spatial infinity.
The relation between the Lagrangian displacement
$\zeta^\alpha$ and the velocity perturbation $\delta \hat{u}^\alpha$ is then given as
an algebraic equation:
\begin{eqnarray}
\delta \hat{u}^\alpha = i \sigma \, e^{-\nu(r)} \,  \zeta^\alpha  .
\label{def-disp}
\end{eqnarray}
Note that the gauge freedom in $\zeta^\alpha$ has been used to demand the
relation $u_\alpha \zeta^\alpha =0$.
The velocity field of a neutron star in a perturbed state may be written as
\begin{eqnarray}
u^\alpha = u_{(0)}^\alpha + \eta\delta \hat{u}^\alpha  \, + \, O(\eta^2) ,
\end{eqnarray}
where $\eta$ is a small expansion parameter introduced for our convenience.
The mass-energy density $\rho$ and the pressure $p$ in the perturbed state
are also given by
\begin{equation}
\rho = \rho_{(0)} + \eta \, (\rho_{(0)}+p_{(0)})\left(
\frac{\Gamma \delta p}{p_{(0)} } - \zeta^\alpha A_\alpha \right)
+ O(\eta^2)\, ,
\end{equation}
\begin{equation}
p = p_{(0)} + \eta \, \delta p + O(\eta^2) \, ,
\end{equation}
where $A_\alpha$ is the relativistic Schwarzschild discriminant defined as
\begin{equation}
A_\alpha =\frac{1}{\rho_{(0)}+p_{(0)}}\, \nabla_\alpha \rho_{(0)} -
 \frac{1}{ \Gamma p_{(0)}}\, \nabla_\alpha p_{(0)} \, ,
\end{equation}
and $\Gamma$ is the adiabatic index defined as
\begin{eqnarray}
\Gamma = \frac{\rho_{(0)}+p_{(0)}}{p_{(0)}}\,
\left(\frac{\partial p_{(0)}}{\partial \rho_{(0)}}\right)_{ad}
 \, .
\end{eqnarray}
To derive equation (16),
we have used the adiabatic condition for the perturbations:
\begin{eqnarray}
\Delta p = \frac{\Gamma \, p_{(0)}}{\rho_{(0)}+p_{(0)}}\,\Delta \rho \, .
\label{ad-rel}
\end{eqnarray}

Assuming spherical symmetry of the background spacetime, we may expand
the perturbations in terms of appropriate tensor spherical harmonic functions.
The Lagrangian displacement, $\zeta^k$ and the pressure
perturbation, $\delta p/(\rho_{(0)}+p_{(0)})$ can be written as
\begin{equation}
\zeta^r = r S_l(r) Y_l^m (\theta,\varphi) \, e^{i \sigma t} \,  ,
\label{xi-r}
\end{equation}
\begin{equation}
\zeta^\theta = \left(
H_l (r) {\partial {Y_l^m (\theta,\varphi)}\over\partial\theta}
- T_{l'} (r) \frac{1}{\sin \theta} \,
{\partial{Y_{l'}^m (\theta,\varphi)}\over\partial\varphi} \right) \,
e^{i \sigma t} \, ,
\label{xi-th}
\end{equation}
\begin{equation}
\zeta^\varphi = {1\over\sin^2\theta}\, \left( H_l (r)
{\partial {Y_l^m (\theta,\varphi)}\over\partial\varphi}
        + T_{l'} (r) \sin\theta\,{\partial {Y_{l'}^m (\theta,\varphi) }
            \over\partial\theta} \right) \, e^{i \sigma t} \, ,
\label{xi-ph}
\end{equation}
\begin{equation}
\frac{\delta p}{\rho_{(0)}+p_{(0)}} = \delta U_l (r) Y_l^m (\theta,\varphi)
\, e^{i \sigma t} \, ,
\label{del-p}
\end{equation}
where $Y_l^m$ is a spherical harmonic function (Regge \& Wheeler 1957;
Thorne 1980).

By substituting equations (\ref{EMT}), (15), (16), and (17),
together with the perturbations defined by equations (21) to (24),
into equations (\ref{ena-eq}) and (\ref{mom-eq}), and collecting the terms proportional to
the small expansion parameter $\eta$, we obtain a system of perturbation
equations for the solid crust:
\begin{eqnarray}
r\,\frac{d z_1}{dr} = -\left( 1 +\frac{2 \alpha_2}{\alpha_3} +U_2 \right)\,
z_1 +\frac{1}{\alpha_3}\, z_2 +\frac{\alpha_2}{\alpha_3}\, l(l+1)\, z_3
\, ,
\label{b-eq-s1}
\end{eqnarray}
\begin{eqnarray}
r\,\frac{d z_2}{dr} &=& \left\{ (-3-U_2+U_1-e^{2\lambda}\, c_1 \bar\sigma^2 )\,
V_1 +\frac{4\alpha_1}{\alpha_3}\,(3\alpha_2+2\alpha_1)\right\} \, z_1 +
\left( V_2 -4\, \frac{\alpha_1}{\alpha_3} \right) \, z_2
\nonumber \\ && +
\left\{ V_1-2 \alpha_1 \, \left( 1 + \frac{2 \alpha_2}{\alpha_3} \right)
\right\}\, l(l+1)\, z_3 + e^{2 \lambda}\, l(l+1)\, z_4 \, ,
\label{b-eq-s2}
\end{eqnarray}
\begin{eqnarray}
r\,\frac{d z_3}{dr} = - e^{2 \lambda}\, z_1+\frac{e^{2 \lambda}}{\alpha_1}\,
z_4 \, ,
\label{b-eq-s3}
\end{eqnarray}
\begin{eqnarray}
r\,\frac{d z_4}{dr} &=& -\left( -V_1+6\Gamma\, \frac{\alpha_1}{\alpha_3}
\right)\, z_1-\frac{\alpha_2}{\alpha_3}\, z_2 \nonumber \\ && - \left\{
c_1 \bar\sigma^2 V_1 +2 \alpha_1-\frac{2 \alpha_1}{\alpha_3}\,
(\alpha_2+\alpha_3)\, l(l+1) \right\}\, z_3 -
(3+U_2-V_2) z_4 \, ,
\label{b-eq-s4}
\end{eqnarray}
\begin{eqnarray}
r\,\frac{d z_5}{dr} = \frac{e^{2 \lambda}}{\alpha_1}\, z_6 \, ,
\label{b-eq-s5}
\end{eqnarray}
\begin{eqnarray}
r\,\frac{d z_6}{dr} = -(3+U_2-V_2)\, z_6 -\left\{
c_1\bar\sigma^2 V_1 - \alpha_1 (l'+1)(l'-2) \right\} \, z_5 \, ,
\label{b-eq-s6}
\end{eqnarray}
where the dependent variables $z_1$ to $z_6$ are defined as
\begin{eqnarray}
z_1 = S_l (r) \, ,
\end{eqnarray}
\begin{eqnarray}
z_2 = 2 \alpha_1 e^{-\lambda} \frac{d}{dr} \left( r e^{\lambda} S_l (r)
\right) + \left(\Gamma - \frac{2}{3}\, \alpha_1 \right) \, \left\{
\frac{e^{-\lambda}}{r^2}\,\frac{d}{dr} \left( r^3 e^{\lambda} S_l (r)
\right) -l(l+1) H_l (r) \right\} \, ,
\end{eqnarray}
\begin{eqnarray}
z_3 = H_l (r) \, ,
\end{eqnarray}
\begin{eqnarray}
z_4 = \alpha_1 \, \left( e^{-2 \lambda} r\,\frac{d H_l(r)}{dr} + S_l(r)
\right) \, ,
\end{eqnarray}
\begin{eqnarray}
z_5 = T_{l'} (r) \, ,
\end{eqnarray}
\begin{eqnarray}
z_6 = \alpha_1 \, e^{-2 \lambda} r\,\frac{d T_{l'} (r)}{dr} \,
\end{eqnarray}
and the various quantities which appear in the coefficients are
\begin{eqnarray}
\alpha_1 = \frac{\mu}{p_{(0)}} \, , \ \ \ \ \ \alpha_2 = \Gamma -
\frac{2}{3}\, \alpha_1 \, ,  \ \ \ \ \ \alpha_3 = \Gamma +
\frac{4}{3}\, \alpha_1 \, ,
\end{eqnarray}
\begin{eqnarray}
V_1 = \left(1+\frac{\rho_{(0)}}{\varepsilon p_{(0)}}\right)\,
r\, \frac{d \nu }{dr} \, , \ \ \ \ \
V_2 = \frac{\rho_{(0)}}{\varepsilon p_{(0)}}\,
r\, \frac{d \nu }{dr} \, , \ \ \ \ \
U_1 =  \left( \frac{d \nu}{dr} \right)^{-1} \,
\frac{d}{dr} \left( r\, \frac{d \nu}{dr} \right) \, , \ \ \ \ \
U_2 = r\, \frac{d \lambda}{dr} \, , \ \ \ \ \
\label{def-fun1}
\end{eqnarray}
\begin{eqnarray}
c_1 = \varepsilon\,\frac{M}{R^3} \, r \, e^{-2 \varepsilon \nu} \,
\left( \frac{d \nu}{dr} \right)^{-1} \,  ,
\end{eqnarray}
\begin{eqnarray}
e^{2 \lambda}=\left(1-\varepsilon\,\frac{2 M(r)}{r}\right)^{-1}  \, , \ \ \ \ \
M(r) = \int^r_0 4\pi r^2 \rho_{(0)} dr   \, ,
\end{eqnarray}
\begin{eqnarray}
r\, \frac{d \nu }{dr} = \varepsilon\,e^{2 \lambda} \,
\left( 4\pi r^2 \varepsilon p_{(0)}+ \frac{M(r)}{r}\right) \, , \ \ \ \ \
 r\, \frac{d \lambda}{dr} = \varepsilon\,e^{2 \lambda} \,
\left( 4\pi r^2 \rho_{(0)} - \frac{M(r)}{r}\right) \, .
\label{def-fun2}
\end{eqnarray}
Here, $M=M(R)$ and $R$ are the mass and the radius of the star,
and $\bar\sigma=\sigma\sqrt{R^3/M}$ is the dimensionless frequency.
In equations (\ref{def-fun1})-(\ref{def-fun2}), we have introduced a factor
$\varepsilon$ to indicate
the hidden factor $1/c^2$ that represents the strength of
the general relativistic degree.
Fully relativistic oscillation
equations for the solid crust are regained for $\varepsilon=1$, and
the Newtonian oscillation equations are obtained
in the limit of $\varepsilon\rightarrow 0$,
which are the same as those derived by McDermott, van Horn, \& Hansen (1988).
We can see that equations
(\ref{b-eq-s1})-(\ref{b-eq-s4}) are decoupled from equations
(\ref{b-eq-s5})-(\ref{b-eq-s6}) because the
background spacetime is spherical symmetric.
The former describe spheroidal (or polar parity)
oscillations, and the latter toroidal (or axial parity) ones.

In the fluid regions, the oscillation equations to be solved are given by
\begin{eqnarray}
r \, \frac{d  y_1}{dr} =
- \left( 3 - \frac{V_1}{\Gamma} + U_2 \right) \, y_1
- \left( \frac{V_1}{\Gamma} - \frac{l(l+1)}{c_1 \bar\sigma^2} \right)
\, y_2 \, ,
\label{b-eq-f1}
\end{eqnarray}
\begin{eqnarray}
r \, \frac{d y_2}{dr} =
( e^{2\lambda}\,c_1 \bar\sigma^2 + r A_r ) \, y_1-(U+r A_r) y_2 \, ,
\label{b-eq-f2}
\end{eqnarray}
where
\begin{eqnarray}
y_1 = S_l (r) \, , \ \ \ \ \
y_2 = \left(r \, \frac{d \nu}{dr} \right)^{-1} \, \delta U_l (r) \,=\, c_1\bar\sigma^2 H_l .
\end{eqnarray}
Note that equations (42) and (43), which come from equations
(\ref{b-eq-s1})-(\ref{b-eq-s4})
in the limit of $\mu\rightarrow 0$, are essentially the same as those
derived by McDermott, van Horn, \& Scholl (1983) for spheroidal oscillation modes.
Equations (\ref{b-eq-s5})-(\ref{b-eq-s6}), which describe toroidal torsional
oscillations, become trivial in the limit of $\mu\rightarrow 0$ because
there is no restoring force that is responsible for the toroidal torsional oscillations.
To calculate the spheroidal modes,
equations (\ref{b-eq-s1})-(\ref{b-eq-s4}) are integrated
in the solid crust, and equations (\ref{b-eq-f1})-
(\ref{b-eq-f2}) in the fluid core and the surface ocean.
For the toroidal torsional modes, equations (\ref{b-eq-s5})-(\ref{b-eq-s6}) should be
solved only in the solid crust.

For the spheroidal modes the outer boundary condition is given
at the stellar surface ($r=R$)
by $\Delta p = 0$, which reduces to
\begin{eqnarray}
y_1 - y_2 = 0 \, ,
\end{eqnarray}
and the inner boundary condition is the regularity condition at the stellar center
given by
\begin{eqnarray}
c_1 \bar\sigma^2 y_1 - l y_2 = 0 \, .
\end{eqnarray}
The jump conditions at the interfaces between the fluid and
the solid regions are given by the continuity conditions,
across the perturbed interface, of the
stress $P_{\alpha}$ on the interface, where
\begin{eqnarray}
P_{\alpha}=(pq_{\alpha\beta}-2 \mu\Sigma_{\alpha\beta})\, N^\beta,
\end{eqnarray}
and $N^\beta$ is the unit vector normal to the perturbed interface
(see, e.g., Finn 1990).
The normal form of the core/crust or crust/ocean
interface is given by $d(r-\eta\,\zeta^r) = n_\alpha dx^\alpha$ for the radial
coordinate of the perturbed interface given
by $r=R_i+\eta\,\zeta^r$,
where $R_i$ is the radius of the core/crust or crust/ocean
interface in the unperturbed state.
Thus, the
corresponding unit normal one-form $N_\alpha$ have the components as
follows:
\begin{eqnarray}
N_t &=& -e^\lambda\, \eta \, i \sigma r S_l(r)  \, Y_l^m (\theta,\varphi) \,
e^{i \sigma t} +O(\eta^2)\, ,
\nonumber \\
N_r &=& e^\lambda  +O(\eta^2)\, , \\
N_\theta  &=& - e^\lambda\, \eta \, r S_l(r) \,
{\partial {Y_l^m (\theta,\varphi)}\over\partial\theta} \, e^{i \sigma t}
+O(\eta^2) \, , \nonumber \\
N_\varphi &=& - e^\lambda\, \eta \, r S_l(r) \,
{\partial {Y_l^m (\theta,\varphi)}\over\partial\varphi} \, e^{i \sigma t}
+O(\eta^2) \, , \nonumber
\end{eqnarray}
The jump conditions for the spheroidal modes at the interface are then given by
\begin{eqnarray}
y_1 = z_1 \, , \ \ \ \ V_1(y_1-y_2) = z_2\, , \ \ \ \ z_4 = 0\, .
\end{eqnarray}
For the toroidal torsional modes, the boundary conditions are applied
at the top and the bottom of the solid crust, and they are
\begin{eqnarray}
z_6 = 0\, .
\label{b-con-ax}
\end{eqnarray}
By imposing the boundary conditions given
above, we can solve our basic equations as an eigenvalue problem with
respect to the eigenvalue $\bar\sigma$. Here, we employ a Henyey type
relaxation method to obtain numerical solutions to our basic equations
(see, e.g., Unno et al. 1989).

\section{Numerical Results}

Neutron star models that we use in this paper are the same as those
used in the modal analysis by McDermott, Van Horn, \& Hansen (1988).
These models are taken from the evolutionary sequences for cooling neutron
stars calculated by Richardson et al. (1982), where the envelope structure
is constructed by following Gudmundsson, Pethick \& Epstein (1983).
They are composed of a fluid core, a solid crust and a surface
fluid ocean.
The interior temperature is finite and is not constant as
a function of the radial distance $r$.
The models are not barotropic and
the Schwarzschild discriminant $\vert A\vert$ has finite values in the
interior of the star.
The models we use are called NS05T7, NS05T8, and NS13T8, and
their physical properties such as the total mass $M$, the
radius $R$, the central density $\rho_c$, the central temperature $T_c$
and the relativistic factor $GM/c^2 R$ are summarized in Table 1 (for other
quantities, see McDermott et al. 1988).

To classify the various oscillation modes of the three component models, we
use almost the same nomenclature as that employed by McDermott et al. (1988).
We let $p_k$ refer to the acoustic modes of the $k$ th overtone.
The internal gravity modes confined in the surface ocean are denoted as
$g_k^s$ and those in the fluid core as $g_k^c$, where $k$ indicates the overtone number.
The eigenfrequencies of the $f$ modes are usually found
between the $p_1$ and $g_1$ modes.
Associated with the fluid-solid interfaces in the models,
there are two interfacial modes, which we denote as $i_1$ and $i_2$ such that
$\sigma(i_1)\le\sigma(i_2)$.
The $s_k$ modes are spheroidal shear dominated modes of the $k$ th overtone,
the amplitudes of which are strongly confined in the solid crust.
The $t_k$ modes are toroidal shear dominated modes of the $k$ th overtone
propagating only in the solid crust.
Note that the $p_k$-, $g_k$-, $f$-, $i_{1(2)}$, and $s_k$ modes are classified as
spheroidal modes while the $t_k$ modes as toroidal modes.

In the Newtonian limit of $\varepsilon\rightarrow 0$,
we calculate various oscillation modes with $l=2$
for the three neutron star models NS05T7, NS05T8, and NS13T8, and tabulate
the Newtonian eigenfrequencies, which we denote as $\bar\sigma_N$,
in Tables 2 through 4.
These tables confirm that
the Newtonian frequencies obtained in this paper
are in good agreement with those computed by McDermott et al. (1988).
Assuming $\varepsilon=1$, we also compute
the corresponding relativistic oscillation modes with $l=2$,
and tabulate the relativistic eigenfrequencies, which we denote as $\bar\sigma_{GR}$,
in Tables 2 through 4, in which the relative frequency differences defined as
$\delta\sigma/\sigma\equiv(\sigma_{GR}-\sigma_N)/\sigma_{GR}$
are also given.
The relative difference $\delta\sigma/\sigma$ takes various values
for different oscillation
modes, and the value of $\delta\sigma/\sigma$ of a given mode is also dependent on
the physical properties of the neutron star models.
We find $|\delta\sigma/\sigma|\ltsim 0.1$ for most of the oscillation modes we calculate.
The elastic $s_k$ and $t_k$ modes with $k\ge 1$ have particularly small values of
$|\delta \sigma/\sigma|$, which are at most a few percent.
However, there are some exceptions, and
the values of $\delta\sigma/\sigma$ become as large as
$-0.45$ for the $t_0$ modes and $\sim -1$ for the $g^s$- and $g^c$-modes for the model NS13T8,
which is the most compact one among the three.
If we consider the gravitational redshift as one of the general relativistic effects
that are responsible for the difference from the Newtonian oscillation frequency, we may have
$\tilde\sigma_{GR}=e^{\nu(r)}\sigma_N$ and hence
$(\tilde\sigma_{GR}-\sigma_N)/\tilde\sigma_{GR}=1-e^{-\nu(r)}$, which is negative
since $\nu(r)<0$.
Although this estimation gives a consistent result for the surface $g^s$ modes confined
in the surface ocean and for the core $g^c$ modes trapped in the fluid core,
it does not always give good estimations of the difference
as suggested by the tables.
In fact, we have positive $\delta\sigma/\sigma$ for some oscillation modes.
This is understandable, however,
since the eigenfunctions usually possess large radial extent in the interior of stars, and
the eigenfrequencies are dependent on the equilibrium quantities which contain terms
and factors which are absent in the Newtonian oscillation equations.

For the various oscillation modes of NS13T8,
we display the displacement vector
$\zeta^\alpha$ versus the radial coordinates
$r/R$ or $\log (1-r/R)$ in Figures 1 through 8, where
the amplitude normalization of the eigenfunctions is given by $y_1=1$ at $r=R$ for
the spheroidal modes and $z_5=1$ at $r=R_{out}$ for the toroidal modes
with $R_{out}$ being the radius of the ocean-crust interface.
Notice that for the $g^c$ modes, the normalization condition $H_l(R_{in}) = 1$ is
adapted
because the modes are very insensitive to the behavior of the eigenfunctions in the
surface ocean, where $R_{in}$ is the radius at the bottom of the crust.
The eigenfunctions obtained in the Newtonian
calculations are also plotted for comparison in each of the figures.
From the figures 1 through 8, we can see that the basic properties of the
eigenfunctions of the relativistic modes are not very much different from those
of the corresponding Newtonian modes, except for the interfacial modes
(see the next paragraph).
In this sense, we can say that
general relativity does not bring about any essential changes in
the modal properties represented by the eigenfunctions.

As shown by Figures 3 and 4, the
relativistic eigenfunctions of the interfacial modes $i_1$ and $i_2$
are at first sight quite different from the Newtonian eigenfunctions.
Since the eigenfunctions of the relativistic $i_{1(2)}$ mode are rather similar to
those of the Newtonian $i_{2(1)}$ mode, it is tempting to make a guess that
the modal properties
have been exchanged between the two modes when we move from Newtonian dynamics to
relativistic dynamics.
In their Newtonian calculation,
McDermott et al. (1988) showed that the properties of the $i$ modes
are very sensitive to the changes in the physical quantities near the interfaces, and
mentioned a numerical experiment in which they computed the two interfacial modes
by artificially reducing the bulk modulus $\mu$ and found an avoided crossing
between the two modes
through which the modal characters are interchanged with each other.
Inspired by this report, we have carried out a numerical experiment,
calculating the two interfacial modes as a function of $\varepsilon$ for the model NS13T8.
Plotting the frequencies of the two modes versus $\varepsilon$ in Figure 9,
we confirm the occurrence of an avoided crossing between the two modes
at $\varepsilon \sim 0.67$,
through which the modal characters of the two have been exchanged with each other.
As $\varepsilon$ increases from $\varepsilon=0$ (Newtonian limit),
the effective local shear modulus $\mu$ in the solid crust decreases
because of the gravitational redshift effects, as a result of which
the two interfacial modes
experience the avoided crossing when we
move from Newtonian dynamics to relativistic dynamics.

\section{Conclusion}

In this paper, we have calculated a variety of oscillation modes
of relativistic neutron stars with a solid crust in the relativistic
Cowling approximation, in which all metric perturbations are ignored.
We find $|\delta\sigma/\sigma|$ is less than
$\sim0.1$ for most of the oscillation modes with $l=2$, although
there are some exceptions such as the surface gravity modes, the core gravity modes,
and the nodeless toroidal torsional modes.
We also find that the essential modal properties represented by the eigenfunctions
are not strongly affected by introducing general relativity in the sense that
the relativistic eigenfunctions of a mode have almost the same properties as the
corresponding Newtonian eigenfunctions.
An exception may be the two interfacial modes whose eigenfunctions differ from
their Newtonian eigenfunctions, and
we have shown that this can be explained as a result of the mode exchange
between the two modes through an avoided crossing that occurs
when we move from Newtonian dynamics to relativistic dynamics.

Needless to say, nonradial oscillations of neutron stars should be treated
within the framework of general relativity because of their strong gravity.
At present, however, we cannot
investigate the pulsations of neutron stars in all the aspects by using fully
general relativistic formalism because of the complexity general relativity
brings about.\footnote{One of the problems that appears in general relativistic 
formalism is the existence of gravitational waves and a necessity to deal with 
a complex eigenvalue problem corresponding to quasi-normal modes. Some authors 
have treated this problem (e.g., Lindblom \& Detweiler 1983; 
Leins, Nollert \& Soffel 1993; Andersson, Kokkotas, \& Schutz 1995).}
One of the simplest ways to study the oscillations of a neutron star is to
treat the problems within the framework of Newtonian dynamics, that is, to
solve the Newtonian oscillation equations for neutron star models constructed with
the Newtonian hydrostatic equations.
However, it is obvious that
this Newtonian treatment cannot be fully approved for neutron stars
since the Newtonian equilibrium models do not give us
correct mass and radius for the stars.
To reduce the distance between fully relativistic and Newtonian
calculations of nonradial oscillations of the stars,
we may use relativistic equilibrium models and solve the Newtonian
oscillation equations, which was the strategy taken by
McDermott, van Horn, \& Hansen (1988).
In this paper, as an extension of the study by McDermott, van Horn, \& Hansen (1988),
we solved relativistic oscillation equations derived in the relativistic
Cowling approximation for relativistic neutron star models with a solid crust.
Our calculation
confirmed that the classification scheme McDermott, van Horn, \& Hansen (1988)
employed for the oscillation modes of neutron stars with a solid crust is valid
even if we integrate relativistic oscillation equations.

\begin{acknowledgements}
S.Y. acknowledges financial support from the Portuguese FCT through a Sapiens project, 
number 36280/99.
\end{acknowledgements}

\newpage

\begin{table*}
\caption{\label{ns-model} Neutron Star Models}
\begin{tabular}{cccccc}
\hline
\hline
Model & $M\ (M_{\sun})$ & $R$ (km) & $\rho_c$ (g $\rm cm^3$) &
 $T_c$ (K) & $GM/(c^2 R)$ \\
\hline
NS05T7&$0.503$&$9.839$&$9.44\times 10^{14}$&$1.03\times 10^7$ &
$7.54\times 10^{-2}$ \\
NS05T8&$0.503$&$9.785$&$9.44\times 10^{14}$&$9.76\times 10^7$ &
$7.59\times 10^{-2}$ \\
NS13T8&$1.326$&$7.853$&$3.63\times 10^{15}$&$1.05\times 10^8$ &
$2.49\times 10^{-1}$ \\
\hline 
\end{tabular}
\end{table*}

\begin{table*}
\caption{\label{fre-ns05t7} Eigenfrequencies $\bar\sigma$ ($l=2$) of NS05T7 }
\begin{tabular}{cllr}
\hline
\hline
mode & $\bar\sigma_N$ & $\bar\sigma_{GR}$ & $\delta\sigma/\sigma$ \\
\hline
$(Spheroidal) $&$ $&$ $&$ $ \\
$g^c_1$&$ 1.998 \times 10^{-5} $&$ 1.625 \times 10^{-5} $&
$ -0.230 $ \\
$g^s_2$&$ 2.037 \times 10^{-3} $&$ 1.760 \times 10^{-3} $&
$ -0.158 $ \\
$g^s_1$&$ 3.230 \times 10^{-3} $&$ 2.787 \times 10^{-3} $&
$ -0.159 $ \\
$i_1  $&$ 8.282 \times 10^{-3} $&$ 7.499 \times 10^{-3} $&
$ -0.104 $ \\
$i_2  $&$ 1.029 \times 10^{-1} $&$ 9.717 \times 10^{-2} $&
$ -0.059 $ \\
$s_1  $&$ 3.093 \times 10^{-1} $&$ 3.159 \times 10^{-1} $&
$  0.021 $ \\
$s_2  $&$ 5.556 \times 10^{-1} $&$ 5.548 \times 10^{-1} $&
$ -0.001 $ \\
$f    $&$ 1.886 \times 10^{ 0} $&$ 1.823 \times 10^{ 0} $&
$ -0.034 $ \\
$p_1  $&$ 4.038 \times 10^{ 0} $&$ 3.978 \times 10^{ 0} $&
$ -0.015 $ \\
$p_2  $&$ 4.736 \times 10^{ 0} $&$ 4.710 \times 10^{ 0} $&
$ -0.006 $ \\
$(Toroidal) $&$ $&$ $&$ $ \\
$t_0  $&$ 4.019 \times 10^{-2} $&$ 3.619 \times 10^{-2} $&
$ -0.111 $ \\
$t_1  $&$ 3.286 \times 10^{-1} $&$ 3.279 \times 10^{-1} $&
$ -0.002 $ \\
$t_2  $&$ 5.633 \times 10^{-1} $&$ 5.609 \times 10^{-1} $&
$ -0.004 $ \\
$t_3  $&$ 7.346 \times 10^{-1} $&$ 7.355 \times 10^{-1} $&
$  0.001 $ \\
\hline 
\end{tabular}
\end{table*}

\begin{table*}
\caption{\label{fre-ns05t8} Eigenfrequencies $\bar\sigma$ ($l=2$) of NS05T8 }
\begin{tabular}{cllr}
\hline
\hline
mode & $\bar\sigma_N$ & $\bar\sigma_{GR}$ & $\delta\sigma/\sigma$ \\
\hline
$(Spheroidal) $&$ $&$ $&$ $ \\
$g^c_1$&$ 2.248 \times 10^{-4} $&$ 1.821 \times 10^{-4} $&
$ -0.234 $ \\
$g^s_2$&$ 1.297 \times 10^{-2} $&$ 1.102 \times 10^{-2} $&
$ -0.177 $ \\
$g^s_1$&$ 1.507 \times 10^{-2} $&$ 1.289 \times 10^{-2} $&
$ -0.169 $ \\
$i_1  $&$ 5.346 \times 10^{-2} $&$ 5.894 \times 10^{-2} $&
$  0.093 $ \\
$i_2  $&$ 1.002 \times 10^{-1} $&$ 9.236 \times 10^{-2} $&
$ -0.085 $ \\
$s_1  $&$ 3.251 \times 10^{-1} $&$ 3.299 \times 10^{-1} $&
$  0.014 $ \\
$s_2  $&$ 5.755 \times 10^{-1} $&$ 5.746 \times 10^{-1} $&
$ -0.002 $ \\
$f    $&$ 1.873 \times 10^{ 0} $&$ 1.811 \times 10^{ 0} $&
$ -0.035 $ \\
$p_1  $&$ 4.102 \times 10^{ 0} $&$ 4.027 \times 10^{ 0} $&
$ -0.019 $ \\
$p_2  $&$ 4.766 \times 10^{ 0} $&$ 4.747 \times 10^{ 0} $&
$ -0.004 $ \\
$(Toroidal) $&$ $&$ $&$ $ \\
$t_0  $&$ 3.999 \times 10^{-2} $&$ 3.600 \times 10^{-2} $&
$ -0.111 $ \\
$t_1  $&$ 3.419 \times 10^{-1} $&$ 3.410 \times 10^{-1} $&
$ -0.003 $ \\
$t_2  $&$ 5.841 \times 10^{-1} $&$ 5.812 \times 10^{-1} $&
$ -0.005 $ \\
$t_3  $&$ 8.148 \times 10^{-1} $&$ 8.128 \times 10^{-1} $&
$ -0.002 $ \\
\hline
\end{tabular}
\end{table*}

\begin{table*}
\caption{\label{fre-ns13t8} Eigenfrequencies $\bar\sigma$ ($l=2$) of NS13T8 }
\begin{tabular}{cllr}
\hline
\hline
mode & $\bar\sigma_N$ & $\bar\sigma_{GR}$ & $\delta\sigma/\sigma$ \\
\hline
$(Spheroidal) $&$ $&$ $&$ $ \\
$g^c_1$&$ 1.269 \times 10^{-4} $&$ 6.017 \times 10^{-5} $&
$ -1.109 $ \\
$g^s_2$&$ 6.115 \times 10^{-3} $&$ 3.126 \times 10^{-3} $&
$ -0.956 $ \\
$g^s_1$&$ 7.548 \times 10^{-3} $&$ 3.905 \times 10^{-3} $&
$ -0.933 $ \\
$i_1  $&$ 1.862 \times 10^{-2} $&$ 2.357 \times 10^{-2} $&
$  0.210 $ \\
$i_2  $&$ 3.326 \times 10^{-2} $&$ 2.985 \times 10^{-2} $&
$ -0.114 $ \\
$s_1  $&$ 4.512 \times 10^{-1} $&$ 4.528 \times 10^{-1} $&
$  0.004 $ \\
$s_2  $&$ 7.704 \times 10^{-1} $&$ 7.685 \times 10^{-1} $&
$ -0.002 $ \\
$f    $&$ 1.434 \times 10^{ 0} $&$ 1.249 \times 10^{ 0} $&
$ -0.148 $ \\
$p_1  $&$ 3.967 \times 10^{ 0} $&$ 3.038 \times 10^{ 0} $&
$ -0.306 $ \\
$p_2  $&$ 5.511 \times 10^{ 0} $&$ 4.602 \times 10^{ 0} $&
$ -0.197 $ \\
$(Toroidal) $&$ $&$ $&$ $ \\
$t_0  $&$ 1.897 \times 10^{-2} $&$ 1.309 \times 10^{-2} $&
$ -0.449 $ \\
$t_1  $&$ 4.534 \times 10^{-1} $&$ 4.533 \times 10^{-1} $&
$ -0.000 $ \\
$t_2  $&$ 7.727 \times 10^{-1} $&$ 7.695 \times 10^{-1} $&
$ -0.004 $ \\
$t_3  $&$ 1.105 \times 10^{ 0} $&$ 1.100 \times 10^{ 0} $&
$ -0.004 $ \\
\hline
\end{tabular}
\end{table*}

\newpage 

\begin{figure}
\centering
\includegraphics[width=8cm]{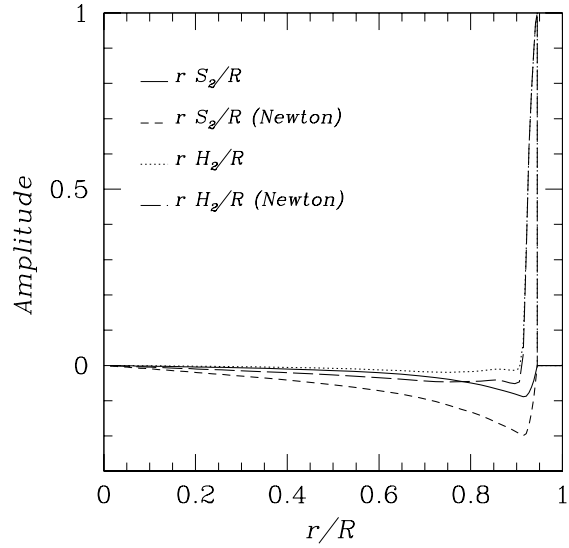}
\caption{
Displacement vectors $r S_2/R$ and $r H_2/R$ of the $g^c_1$ mode
for the model NS13T8, given as a function of $r/R$. Here, the normalization
of the eigenfunction is chosen as $H_2(R_{in})=1$, where $R_{in}$ is the
radius at the bottom of the crust. The Newtonian
eigenfunctions are shown as well as the relativistic ones.
}
\end{figure}

\begin{figure}
\centering
\includegraphics[width=8cm]{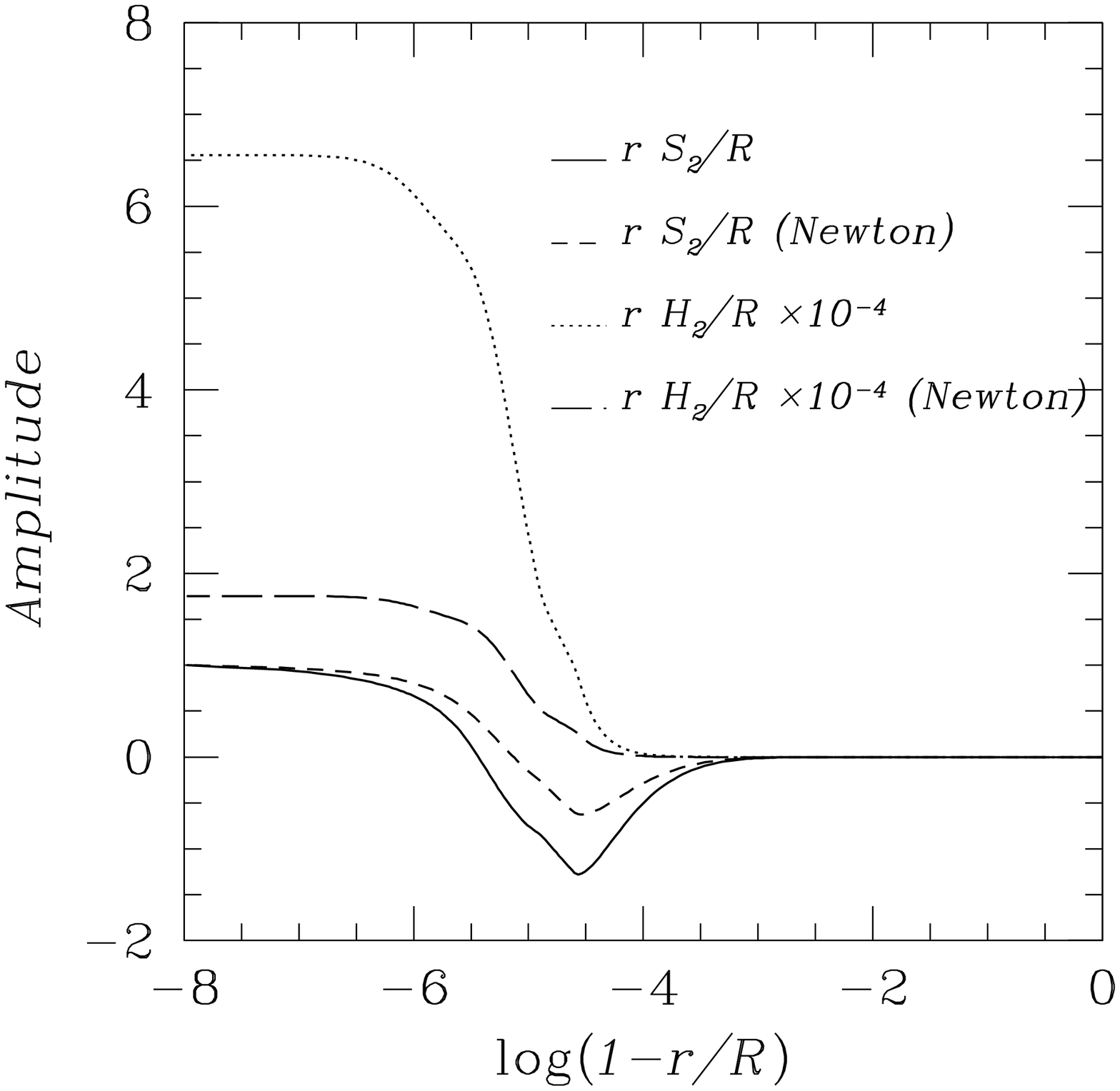}
\caption{
Displacement vectors $r S_2/R$ and $r H_2/R$ of the $g^s_1$ mode
for the model NS13T8, given as a function of $\log (1-r/R)$. Here, the
normalization of the eigenfunction is chosen as $S_2(R)=1$. The Newtonian
eigenfunctions are shown as well as the relativistic ones. 
}
\end{figure}

\begin{figure}
\centering
\includegraphics[width=8cm]{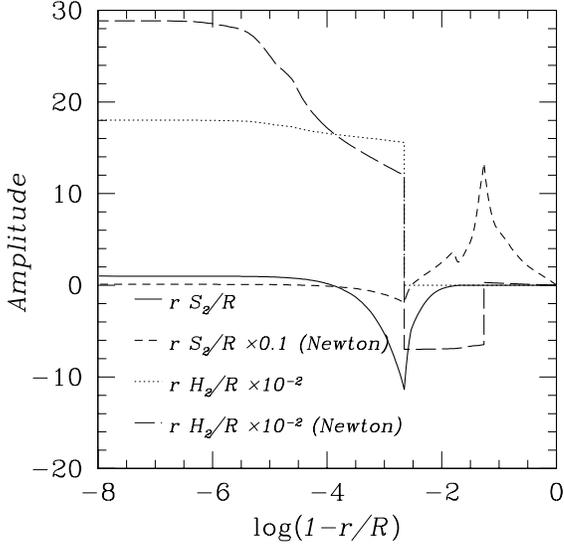}
\caption{
Same as Figure 2 but for the $i_1$ mode.
}
\end{figure}

\begin{figure}
\centering
\includegraphics[width=8cm]{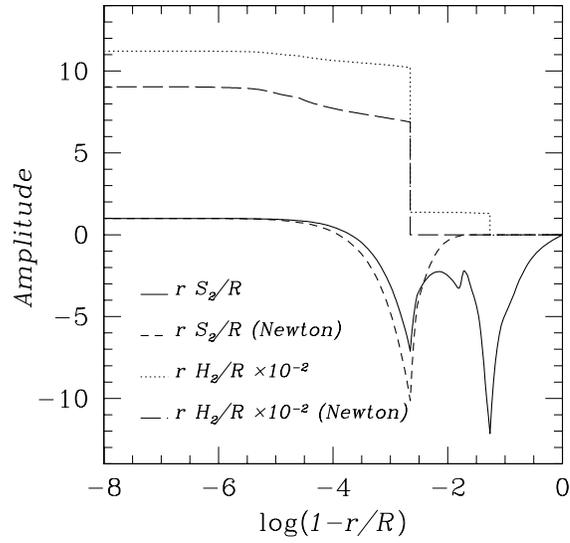}
\caption{
Same as Figure 3 but for the $i_2$ mode.
}
\end{figure}

\begin{figure}
\centering
\includegraphics[width=8cm]{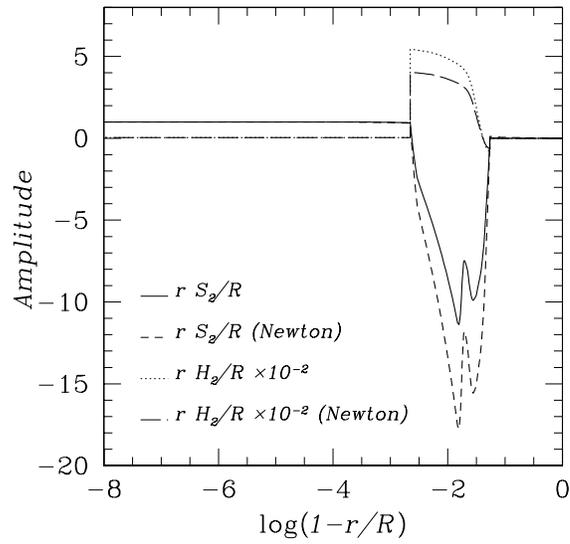}
\caption{
Same as Figure 4 but for the $s_1$ mode.
}
\end{figure}

\begin{figure}
\centering
\includegraphics[width=8cm]{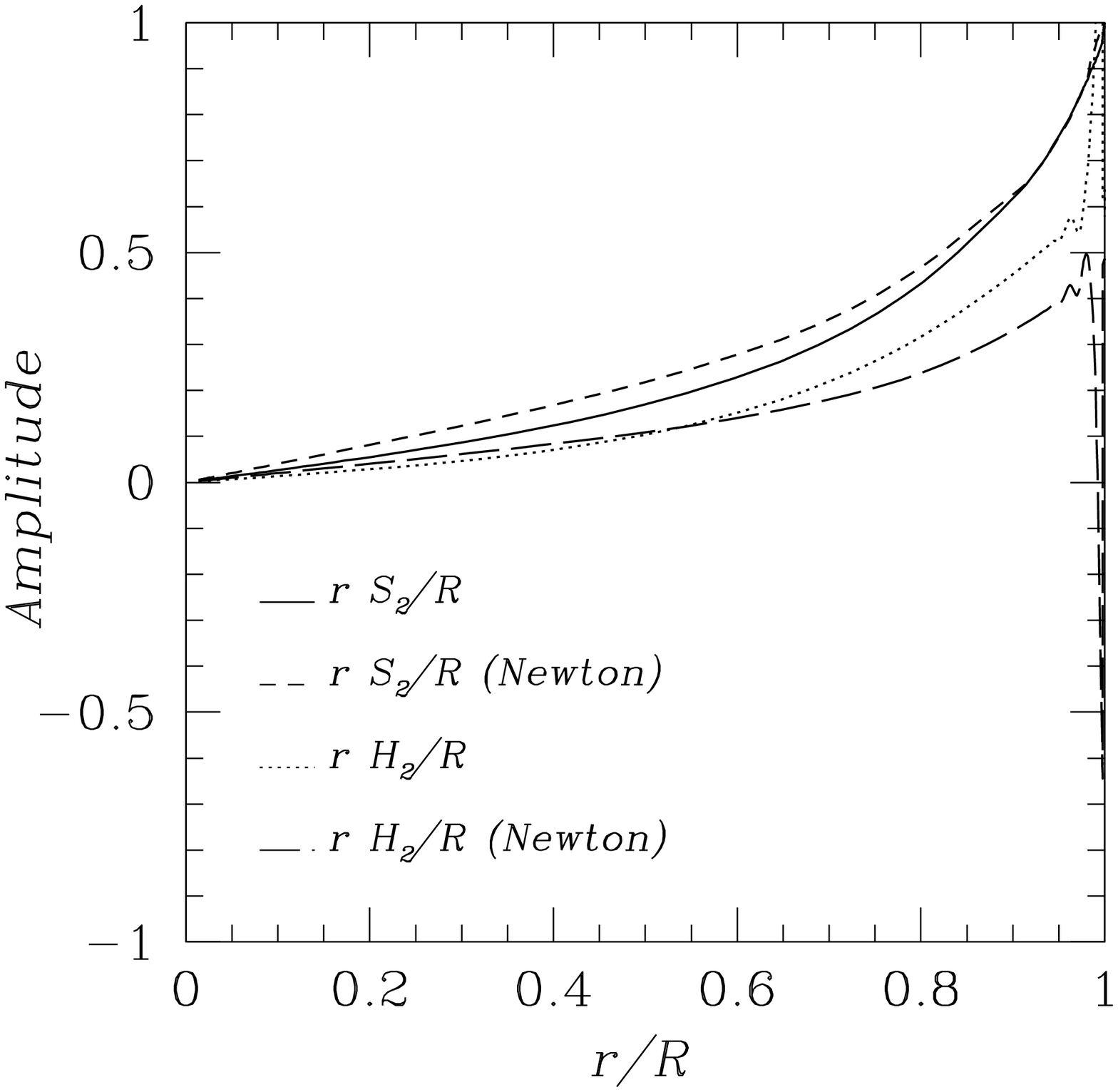}
\caption{
Displacement vectors $r S_2/R$ and $r H_2/R$ of the $f$ mode
for the model NS13T8, given as a function of $r/R$. Here, the normalization
of the eigenfunction is chosen as $S_2(R)=1$. The Newtonian
eigenfunctions are shown as well as the relativistic ones.
}
\end{figure}

\begin{figure}
\centering
\includegraphics[width=8cm]{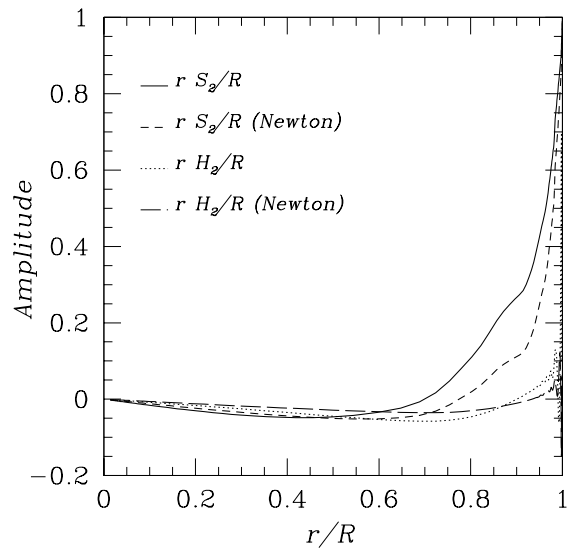}
\caption{
Same as Figure 6 but for the $p_1$ mode.
}
\end{figure}

\begin{figure}
\centering
\includegraphics[width=8cm]{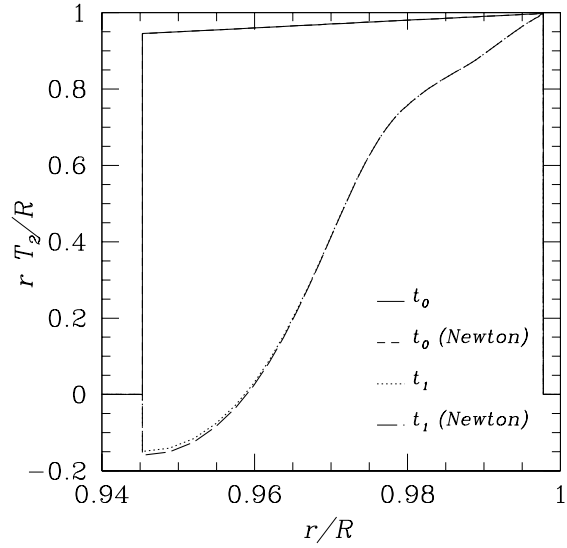}
\caption{
Displacement vector $r T_2/R$ of the $t_0$ and the $t_1$ modes
of the neutron star model NS13T8, given as a function of $r/R$. Here,
normalization of the eigenfunction is chosen as $ T_2(R_{out})=1$, where
$R_{out}$ is the radius at the crust/ocean interface. The Newtonian
eigenfunctions are shown as well as the relativistic ones.
Note that two curves for $r T_2/R$ nearly overlap each other.
}
\end{figure}

\begin{figure}
\centering
\includegraphics[width=8cm]{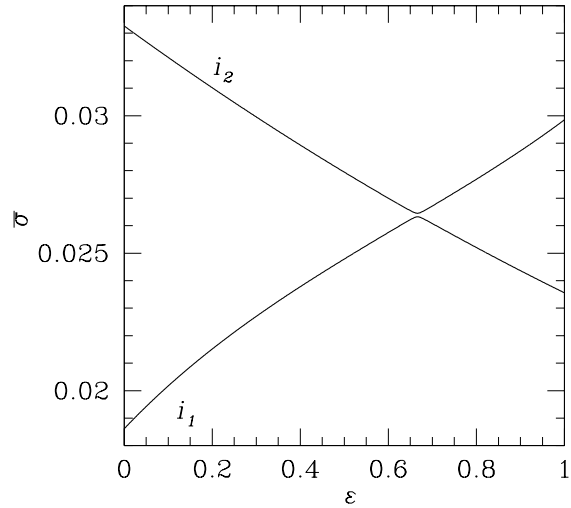}
\caption{
The avoided crossing between the $i_1$ and $i_2$ mode of
the model NS13T8, where $\varepsilon$ is the parameter that represents
the strength of general relativistic effects. The avoided crossing happens
at $\varepsilon \sim 0.67$.
}
\end{figure}

\end{document}